\documentclass[journal]{journal}
\ifCLASSINFOpdf
\else
\fi

\begin{document}
%
\title{Chat-GPT: An AI Based Educational Revolution}
%

\author{
    \IEEEauthorblockN{ Sasa Maric}
    , \IEEEauthorblockN{Sonja Maric}
    , \IEEEauthorblockN{Lana Maric}
    \\
    \IEEEauthorblockA{\textit{Uiveristy of NSW, Sydney, NSW, Australia}}\\
    \IEEEauthorblockA{\textit{Department of Education, Sydney, NSW, Australia}}\\
    \IEEEauthorblockA{\textit{School of Education, Arts and Social Sciences Macquarie University, Sydney, NSW, Australia}}
}
\markboth{Journal of \LaTeX\ Class Files,~Vol.~6, No.~1, January~2007}%
{Shell \MakeLowercase{\textit{et al.}}:Chat GPT: An Educational Revolution}

\maketitle
\thispagestyle{empty}

\begin{abstract}
The AI revolution is gathering momentum at an unprecedented rate. Over the past decade, we have witnessed a seemingly inevitable integration of AI in every facet of our lives. Much has been written about the potential revolutionary impact of AI in education. AI has the potential to completely revolutionise the educational landscape as we could see entire courses and degrees developed by programs such as ChatGPT. AI has the potential to develop courses, set assignments, grade and provide feedback to students much faster than a team of teachers. In addition, because of its dynamic nature, it has the potential to continuously improve its content. In certain fields such as computer science, where technology is continuously evolving, AI based applications can provide dynamically changing, relevant material to students. 
AI has the potential to replace entire degrees and may challenge the concept of higher education institutions. We could also see entire new disciplines emerge as a consequence of AI.
This paper examines the practical impact of ChatGPT and why it is believed that its implementation is a critical step towards a new era of education. We investigate the impact that ChatGPT will have on learning, problem solving skills and cognitive ability of students. We examine the positives, negatives and many other aspects of AI and its applications throughout this paper. 
\end{abstract}


%
\IEEEpeerreviewmaketitle

\section{Introduction}
\IEEEPARstart{T}{he} introduction of ChatGPT in late 2022 has provided the general population with an incredible opportunity to take advantage of AI. ChatGPT is an AI based chat bot developed by OpenAI, it was built on top of their already established GPT-3 family. Despite its recent inception into mainstream society, ChatGPT has already caused a stir in higher education, as educators attempt to navigate this new landscape. Many see its introduction as 'game changing' technology that will help revolutionise current educational methods, which are seen as severely outdated. \cite{ed_failure}\cite{ed_failure1}\cite{higher_education_tech}. 

In this paper we look at the advantages that AI based applications such as ChatGPT have over traditional teaching methods. Unlike traditional learning methodologies which are often non malleable, ChatGPT provides a dynamic way for students to learn. ChatGPT has many advantages over traditional learning pedagogies. This paper also discusses some of the positive aspects of AI, but it also analyses the potential negatives of integrating AI into higher education. The negative effects of AI on students learning and their academic performance has largely gone unanswered \cite{dis_AI}. In this paper our objective is to critically analyse not only the general AI trend in education but also the practical implications of an advanced AI technology being implemented into higher education. We consider the impact that AI based applications will have on teaching pedagogy in higher education, we look at how higher education teaching must adapt and propose some simple techniques that can be used by teachers to integrate AI positively into their teaching \cite{education3} \cite{education2}\cite{education1}. 

Modular integration of solutions (analogous to jigsaw puzzles) is a critical for achieving positive outcomes using chat-based application such as ChatGPT. This is particularly true in the analysis of new and novel problems. Students should be encouraged to use solutions to smaller, well known problems that are easily produced by AI application software to produce solutions to harder, more complex problems. AIs flexibility and ability to develop solutions quickly and accurately can facilitate an iteration based approach where solutions are put together and tested frequently. This black box approach where the integration of elements is more important than the internal components of individual elements has the potential to become an entire discipline in many areas of study. 

ChatGPT was developed in late 2022, it was trained using two standard AI techniques. Reinforced learning, which is a action/reward framework where the AI algorithm initially takes random actions and is rewarded according to whether its actions produce positive or negative results. As the number of actions increases, the algorithm is able to optimise its performance and  maximize its reward. ChatGPT was trained using a variety of sources including books, articles and websites indexed from all over the internet. This data was then used to develop and build a unique language model that enables ChatGPT to understand natural language, which helps users interact with the interface. It also allows ChatGPT to generate responses that resemble human language, making its responses easy to read and understand. Supervised learning plays a critical role in how ChatGPT operates. The AI algorithm learns from data that has been labelled by humans. This allows the algorithm to be able to check its calculations and optimise its performance.
 
ChatGPT is a web based application that uses a chat interface to interact with users. Users are able to ask the application questions in a variety of forms. To get the best possible results users should ask detailed questions. The more detailed the question the better the answer. ChatGPT differs from other already available AI based tools such as Siri because of its comprehensive library and its ability to output sophisticated material to its users. For example, teachers are able to develop entire curriculum programs for their classes within seconds, even design practical lessons in physical education classes. Software engineers are able to ask for code for complicated applications. It has been received so well that many of the biggest technological companies are promising rival AI based apps. For example, Google is on the verge of releasing a rival AI chatbot called Bard. Many of the other tech giants will no doubt follow suit.

AI technology is destined to revolutionise how we gather, process and interact with data. It is set to change the way that businesses work and how they offer services. It is also tipped to revolutionise education at every level. There are many positives that AI can offer to both students and teachers. We present a detailed analysis of ChatGPTs positives and negatives in Section II. 

With a move away from traditional education, students must adapt and learn in different ways. Many of the methods for teaching and learning  must be tailored to take into account applications such as ChatGPT. Teachers are forced to implement new teaching pedagogies in order to incorporate AI effectively into their subjects. Assessment and testing of students at tertiary level will also have to be substantially revolutionised, taking into account essay-writing technology like ChatGTP. Traditional testing and assessment creation relied heavily on students having static resources to learn from. The emphasis of assessments was for students to develop an understanding of a topic using research methodologies. Since these questions can now be answered instantly using applications such as ChatGTP we must now look to new ways of assessing students. Section III presents an overview of some potential teaching methodologies that can be applied to best integrate AI into higher education degrees and units. Section IV presents a simple overview of some future implications of AI and Section V concludes the paper.


\section{Advantages and Disadvantages of ChatGPT}
Since its inception, AI has been the subject of much debate. Despite a significant investment into the development of AI in the 60s and 70s, its implementation and integration was delayed due to a number of reasons. Chief among these was a lack of computing power. As computers developed and became more capable, AI became more prevalent. One of the first practical implementations of AI was a program called Eliza \cite{eliza1}. Eliza was the first example of a chat-based AI application. It contained a small database and could only offer superficial responses to questions. But, in many ways applications like Eliza paved the way for programs such as ChatGTP. Ever since Eliza's implementation, a number of ethical considerations have arisen pertaining to the future development of AI. Public opinion is divided on whether research on AI should continue.
\subsection{ChatGPT Advantages}
There are many positive elements of implementing AI into education. We have already seen education revolutionised by rapidly advancing technology that has been developed over the past couple of decades. For example, it is now standard practise for higher education facilitators to offer online learning to students who are not physically able to attend a lessons. Many students have obtained entire degrees without ever setting foot on an university campus. With the implementation of ChatGPT we have reached another defining moment in an ever evolving educational revolution. There is concern that ChatGPT threatens to make higher education institutions completely irrelevant because of its unprecedented ability to gather information and present it in an interactive manner. We will explore this idea in more detail in the sections that follow. 

Previous attempts at developing chat-based AI applications have been rudimentary when compared to ChatGPT. ChatGPT has an extraordinarily extensive database which it is able leverage to give clients detail responses, even with complex queries. One of ChatGPTs advantages over standard online learning material such as video recordings, is that students have the ability to ask questions in a customised way. Unlike with online material (videos and animations), ChatGPT is able to continuously evolve its answers as it is exposed to up to date materials. Students are able to ask the interface questions in a personal and customised way that reflects their level of understanding on a subject. This personalisation of questions is an essential advantage that ChatGPT has over traditional learning methods. 

Using AI chat based applications, students are able to ask questions much like they would in a traditional classroom setting, but in a private and customised manner. Traditional online and face-to -face pedagogy resources provide static reference material. Whereas, ChatGPT is able to provide students with a dynamic and highly personalised way to learn and digest information. It has many advantages over teachers/lecturers as well. A critical advantage that ChatGPT has over teachers, is that is can operate continuously and can simultaneously interact with thousands of students. While a teacher might take days to answer 1000 personalised questions, ChatGPT can develop detailed and relevant answers in a fraction of the time. AI based systems run continuously so they can be accessed any time by students. This opens up opportunities for people that have  work and other obligations. It provides them with extraordinary flexibility. This is further amplified by the extraordinary expertise that is offered by AI based applications. Current part-time students can only gain direct access to teachers few hours per week in set time slots, often either in lectures or in predefined consultation times. With ChatGPT this level of expertise can be accessed anytime. ChatGPT can simultaneously differentiate material for students with different capabilities. This is something that is extremely difficult to achieve with a single teacher. The current method for teaching exposes students to their teachers for a few hours per week and the students often out-number the teacher by 500 to 1. It would be impossible for a teacher to develop a differentiated curriculum for each individual student. AI based applications are able to develop highly personalised units and set curriculum to suit each students learning style and ability. 

Another advantage that AI has over the standard teaching model is AIs ability to keep up with technology. AI is able to continuously update its database to include the most relevant material available. It is almost impossible for a teacher to be able to do that consistently. This is critical in STEM related fields where the landscape changes rapidly. 

With AI Lessons can be developed with an emphasis on concepts and not only on students memorising content. This allows students to develop a greater understanding of the concepts that underlay the content and promotes higher order thinking. This further extends to students after they graduate, it is no longer necessary to have students memorise entire text books, formulas and complicated facts. Application like ChatGTP will provide businesses with the expertise they do not currently have. Students with the ability use ChatGPT and implement its solutions in a tailored manner, are going to be highly sought after.
\subsection{ChatGPT Disadvantages}
AI based applications such as ChatGPT will revolutionise many areas of our businesses and personal lives. In particular, there are a number of significant positives that AI will bring to education in particular. However, much like every other  significant technological advancement there are also a number of negatives that must be analysed to develop a fair and balanced view.

With a move away from direct, traditional teaching, students may miss out on face-to-face learning where they continuously interact with teachers and their fellow students. Constructivist theories of learning suggest that learning is a vastly social process, whereby students learn from each other as well as their educator \cite{construct}. For this reason, there is a danger that students will fall behind and fail to absorb information that is provided swiftly and easily through ChatGPT. Without a teacher or expert facilitating and scaffolding new information, students may be producing work that they do not fully comprehend. Literature on social learning has emphasised the significance of students' interaction with each other, as well as their teachers, and the positive impact is has on their well being and academic performance (ericson reference ). Students who have positive relationships with their teachers are more likely to engage in the content and academic perfomance is increased with positive social environments (Ericson reference) With the inception of AI technologies like ChatGTP, these processes are further threatened \cite{Eric}. 

Much debate has been conducted over the attention span of current students. A great deal of research has been conducted on the effects of social media on the attention spans of children and young adults\cite{attention4} \cite{attention2}. AI threatens to have a negative impact on  attention span as it produces answers to questions almost instantaneously. These instantaneous solution are detrimental because they do not reflect the amount of time and research that must be conducted to get such detailed solutions. This could cause issues for students when they must independently solve novel problems that are not included in the AI based application.

Literature on COVID19 has already found that student social interactions, as a result of isolation, has suffered and their engagement in online content has been inconsistent (studente reference). The integration of AI and technology into student learning can have serious and detrimental implications on social integration in school life. This can be especially true during transitional times in students' lives, like moving from primary school to high school, high school to tertiary education, etc. Having students increasingly dependent on online sources such as Khan Academy, Youtube and now ChatGPT will have strong implications in their social development as they look to move away from an academic environment. A decrease in social interaction with peers, as well as with teachers and lecturers can have a follow-on effect in how these students deal and interact with coworkers, stakeholders and their supervisors (Rogers reference). These effects have the potential to follow them into the future and impact all social aspects of their lives \cite{Studente}.

Developing problem solving skills has always been a high priority for teachers at all stages of education. Many theories have been proposed to gain positive results and an increase students problem solving abilities. In mathematics for example, problems solving skills are developed through the use of simple algorithms along with repeated exposure to problems. It is hoped that continuous exposure to problems will develop procedural awareness to help students solve more complicated problems in the future. Much of the STEM based units focus on breaking down a problem into smaller, more manageable problems which can be tackled one by one, building up the final solution. Their problem solving methodologies, like many other, are difficult to implement with the current stock of AI based systems. However, with future iterations we could see AI based apps develop and implement many of these methodologies to help progress a students problem solving abilities.

This proposes an interesting research question. Could AI based tools help develop problem solving skills? Will it help people with fast and accurate feedback? Or will it have a detrimental impact on the problem solving skills of the new generation of students? An argument can be made for both sides. Reliable, fast and accurate feedback are some of the positives of AI based chat applications such as ChatGPT. However, the fact that students are able to ask and receive answers very quickly, could have a negative impact on how students approach questions. For example, breaking down the problem into smaller easier problems as discussed above would not be beneficial. There is no need to follow procedures or algorithms to arrive at answers. The AI simply allows students to arrive at the answer. However, it can be argued that AI can be developed to incorporate these methods of problem solving.

There is also potential that students creativity could be  impacted with the introduction of AI based tools. For example, student would no longer have to think of innovative ways to solve problems. They can simply use AI applications to answer them, will little or no background knowledge. This is especially problematic when novel problems are presented that are not part of the AI database. Since students do not have the prerequisite knowledge and skills in solving simple problems. It would become difficult for them to develop novel approaches to solve problems. It would also be difficult to gauge the difficulty of problems that are presented to students, because the answers are often easily derived and presented by the AI. Students could have a skewed perception of the difficulty of problems. This is especially important in higher education that is centered around providing students with the core skills that they can use in the workforce. The STEM fields would be particularly effective because much of what the students learn is obsolete even before they graduate. As such, the emphasis on "soft" skills becomes a key consideration and is something that is difficult to develop using AI based applications. Many higher education institutions focus on developing students research, verbal, written and problem solving skills. Since these skills are transferable and can be applied to make applications, they are seen as a key deliverable, of most degree programs, in higher education environments.

\section{Teaching Revolution}
The introduction of ChatGPT and other AI based platforms can be seen as a catalyst for a teaching revolution, particularly in higher education which has been static for many decades. There has been much research based on how online learning and technology can be applied in higher education institutions. The introduction of AI technology will bring about a revolution that will forced a move toward a new teaching pedagogy. As AI develops and gains traction in higher learning institutions, there must be conscious shift toward a new way of teaching.
The structuring of degrees and units will have to be modified to reflect the fact that AI is now available to students. If this technology is available to students to solve problems quickly and accurately, do we look at developing units that use these solutions as black boxes? It could be the case that a modular approach to teaching should be implemented, instead of a technical problem solving one. Where students are assessed on their ability to ask questions and use the solutions to smaller problems, to solve more complex problems. This is particularly evident in STEM subjects, where a technical foundation is desirable. For example, programming courses traditionally teach simple coding which students use to develop simple programs such as calculators. A scaffolding approach is used, where students use these simple concepts to learn more complicated concepts which allow them to solve more complicated problems and produce more complex applications. 

This type of approach can be generalised using AI, but would it be useful for students to understand the underlying work of applications? It might be case that students should instead be taught how to incorporate these programs and treat them as black boxes connected together to achieve a goal. The introduction of AI could force the start of a whole new discipline which would be centered around incorporation of programs much like a jigsaw. 

Methods of assessing students must also be analysed in detail. There are a number conflicting methodologies that have to be considered when looking at what type of assessments students should be subject to in higher education environments. Of course the choice of assessment is dependent on the type of course that is being offered to the student. There is an argument to say that traditional tests are useless and all about remembering facts, which is not how students are going to do work when they become part of the workforce. Take home assessments are often looked at as more favorable because they tests a students ability to solve more complex problems with research and collaboration with other students, which often mirrors their working environment. However, with the availability of application such as ChatGPT, assessments could be easily completed without any research or problem solving ability. One could also make an argument that students would have access these applications once they enter the workplace, so they will be able to use these to solve problems? Should it matter how the problems are solved? What happens when a problem arises that is not part of ChatGPT database? When novel problems arise would students be equipped to solve these problems?

There are many issues that need to investigated and analysed in the context of teaching techniques. How do we encourage students to learn and develop their problem solving skills? How do we set tasks and assessments to foster students skills and develop their understanding of the core concepts of that subject? Teaching and learning is going through a revolutionary phase, where the traditional ideas on education will be challenged. How AI is integrated in this educational revolution is going to be key in developing capable students in the future \cite{Roger}.

\section{Future of ChatGPT}
The future of AI based applications such as ChatGPT, is difficult to predict because of their infinite potential. Most of the current generation of workers and students did not grow up with AI based tools.  As such their dependence on AI based applications is still low. The real potential of such applications is going to be shown with the next generation of students who are going to grow up with this technology as key aspect in their schooling. It is predicted that AI will have a similar impact that the internet had on current generation of students who grew up with the internet as part of their every day lives. 
As discussed briefly in the previous sections AI has the potential to develop and deliver entire degrees. AI has the ability not only to develop the degrees, but to dynamically update these degrees and units as technological advancements are developed. AIs dynamic nature is one of its main advantages over traditional educational frameworks. Applications such as ChatGPT have the potential to replace entire universities and provide students with the very best teaching material and methodologies. 
AI has the potential to be able to develop assignments, tests and weekly exercises. Providing almost immediate feedback to students. Tailoring the way teaching and learning is done. This has the biggest impact in places where higher education is hard to access. AI has the ability to provide the very best education anywhere around the world in a flexible manner. Students can work or look after kids and take lesson when they can. The material can be delivered in a number of different ways and can be tailored to each individual student. This is in contrast to traditional ways of teaching. For example, current higher education units often provide students with a few lectures and some tutorial or practical to help reinforce the concepts that are discussed in the lectures. This is a very static model and a one size fits all model. This is not an optimal situation for many students that may prefer to learn in different ways. For example, some students have a short attention span or prefer to learn with hands on material. It is very difficult to provide this level of differentiation for each student, especially when the cohort of students is large. However, this would be easily accomplished with AI based tools. AI can simultaneously manage and tailor learning material for students to provide optimal results.

ChatGPT is one of many applications that is being developed by organisations around the world. There are many potential benefits of such applications and these extend far beyond education. The applications of such AI based systems is only beginning to be discovered and the future iterations of these AI systems promises to revolutionise the way we do business, access entertainment, build and develop projects.

Another potential implication of AI based applications can be the introduction of new areas of research and study. We could see the emergence of a set of entirely new disciplines where the goal is not to solve problems but to take already solved problems and assemble these into solutions for much bigger problems. A modular engineering or computer science degree where the goal would be developing skills to interact with AI technology, having the ability to ask specific and relevant questions. Then taking the relevant outputs and developing new and novel solutions to problems using a modular approach analogous to solving a jigsaw puzzle.
\section{Conclusion}
ChatGPT is one of the first publicly available AI based applications that has peaked public interest. It has a large database that can be accessed using a chat interface. It is able to answer complex questions and queries in a large variety of subjects. Its implementation is tipped to have vast impact on many industries including education. This paper has introduced some key considerations of such application within the educational framework. We present various advantages and disadvantages of not only ChatGPT but AI based systems in general. We present some consideration that need to be taken into account within higher education. We looked at some possible positives and how AI based technologies can help students learn and develop their skills. How AI based technology can implement flexible, highly personalised courses and degrees for students. Currently, the education sectors application of AI technology is limited. We believe that this is set to change and that AI technology has the potential to replace much of the current educational structure. The future of AI within education is difficult to predict but it is certain that there will be a significant integration of AI in the sector in the near future.


%

\ifCLASSOPTIONcaptionsoff
  \newpage
\fi

\end{document}